\documentclass[preprint,showpacs,prl]{revtex4}
\usepackage{stmaryrd}
\usepackage{amsmath}
\usepackage{amssymb}
\usepackage{graphicx}
\usepackage{dcolumn}
\usepackage{bm}

\begin{document}

\begin{titlepage}

\title{Dirac Fermion in Strongly-Bound Graphene Systems}

\author{Yuanchang Li,$^1$ Pengcheng Chen,$^1$ Gang Zhou,$^1$ Jia Li,$^2$ Jian Wu,$^1$ Bing-Lin Gu,$^3$ S. B. Zhang,$^4$\footnote{zhangs9@rpi.edu} and Wenhui Duan$^{1,3}$\footnote{dwh@phys.tsinghua.edu.cn}}
\address{$^1$Department of Physics and State Key Laboratory of Low-Dimensional Quantum Physics, Tsinghua University, Beijing 100084, People's Republic of China \\
$^2$Institute of Advanced Materials, Graduate School at Shenzhen, Tsinghua University, Shenzhen 518055, People's Republic of China \\
$^3$Institute for Advanced Study, Tsinghua University, Beijing 100084, People's Republic of China \\
$^4$Department of Physics, Applied Physics, and Astronomy,
Rensselaer Polytechnic Institute, Troy, New York 12180, USA}

\date{\today}

\begin{abstract}
It is highly desirable to integrate graphene into existing semiconductor technology, where the combined system is thermodynamically stable yet maintain a Dirac cone at the Fermi level. First-principles calculations reveal that a certain transition metal (TM) intercalated graphene/SiC(0001), such as the strongly-bound graphene/intercalated-Mn/SiC, could be such a system. Different from free-standing graphene, the hybridization between graphene and Mn/SiC leads to the formation of a dispersive Dirac cone of primarily TM $d$ characters. The corresponding Dirac spectrum is still isotropic, and the transport behavior is nearly identical to that of free-standing graphene for a bias as large as 0.6 V, except that the Fermi velocity is half that of graphene. A simple model Hamiltonian is developed to qualitatively account for the physics of the transfer of the Dirac cone from a dispersive system (e.g., graphene) to an originally non-dispersive system (e.g., TM).
\end{abstract}


\maketitle

\draft

\vspace{2mm}

\end{titlepage}
Graphene, a one-atom-thick carbon sheet, has sparked enormous interest during the past few years. Its peculiar band structure, with two linear dispersion bands crossing at the Fermi level ($E_F$) and other outstanding physical properties, make graphene an appealing system for both fundamental studies and modern technological applications \cite{RMP,status,rise,Zuanyi}. Due to the two-dimensional
nature of graphene, however, a substrate is required for most practical applications. The coupling with the substrate, however, leads to modification of the graphene morphology in the form of lattice corrugation, which could significantly affect its electronic properties \cite{Giovannetti,JPCM,Ihm,Varchon,Mattausch}. It is generally believed that to preserve the Dirac cone, graphene should only bind weakly to the substrate via, for example, van der Waals interactions. From a technological point of view, however, the stronger the binding, the better the thermal and mechanical stabilities of the combined system. It is, therefore, imperative to find Dirac cone structures that can also exist in strongly-bound systems with significantly larger binding energy.

In searching for a solution, two issues must be addressed: (i) how to increase the binding without significantly increasing the corrugation of the graphene. For example, graphene/SiC (G/SiC) can be considered as a strongly-bound system due to the formation of two C-Si bonds out of eight carbon atoms per 2 $\times$ 2 graphene cell \cite{Mattausch}. However, because only 25\% of the carbon atoms are involved in such a bonding, the net result is a heightened corrugation of the graphene. In this regard, a buffer layer with less directional bonding between graphene and SiC is highly desirable. (ii) A chemically active buffer layer may cause considerable charge transfer between graphene and substrate. Because graphene is strongly electronegative, most likely its Dirac cone will be filled with electrons. One therefore needs to engineer a new Dirac cone in the vicinity to replace that of graphene; in other words, to transfer the Dirac fermion of graphene to the surface of the substrate. For this to work, clearly, the surface of the substrate must possess at least doubly degenerate states: for example, $p$ states at the symmetry points of the surface Brillouin zone, or transition metal (TM) $d$ states, or rare-earth metal $f$ states.

Using first-principles calculations, we have systematically studied the graphene/intercalated TM/SiC(0001) systems (denoted as G/i-TM/SiC) with 3$d$ TM elements. We find clearly-defined Dirac cone structures centered at the $K$ point of the Brillouin zone as a consequence of the interaction among graphene, SiC $s$, $p$ states and degenerate TM $d$ states. In contrast to weakly-bound systems, Dirac states are primarily located in the TM layer with diverse electronic properties. For example, G/i-Mn/SiC is a non-magnetic system with the Dirac point exactly at $E_F$; G/i-Cr/SiC is a spin-polarized half Dirac fermion system in the sense that there is a Dirac point for majority-spin states 0.2 eV above $E_F$ but not for minority-spin states; G/i-Co/SiC and G/i-Ni/SiC show a gap at the Dirac point due to symmetry lowering, yet both maintain a good linear energy dispersion. A band coupling model is constructed to qualitatively explain the transfer of the Dirac fermion from graphene to TM in these coupled systems. The significant hybridization between graphene, SiC, and TM also results in noticeably stronger binding of graphene to the substrate with a binding energy $E_b$ as high as 0.51 eV/C for G/i-Mn/SiC. Transport calculation reveals that the Dirac fermion of
G/i-Mn/SiC behaves almost the same way as that of free-standing graphene, signifying the high potential of the G/i-TM/SiC systems in future electronic and spintronic applications.

Our electronic structure calculations are performed within the local spin density approximation (LSDA) \cite{CA} as implemented in the Vienna \emph{ab-initio} simulation package (VASP) \cite{vasp}. The electron-ion interaction is described by a projector augmented wave method \cite{PAW} with an energy cutoff of 400 eV. Integration over the Brillouin zone is carried out using the Monkhorst-Pack scheme with 12 $\times$ 12 $\times$ 1 $k$-points. A $\sqrt3 \times \sqrt3R30^{\circ}$ supercell is used for 6$H$-SiC, which can accommodate a 2 $\times$ 2 graphene cell, with one TM atom intercalated between the graphene and SiC substrate (see Fig. 1). The SiC substrate is modeled by a six SiC bilayer slab with hydrogen passivation of the bottom surface. In the simulation, the three bottommost SiC bilayers are fixed at their respective bulk positions whereas all other atomic positions are fully relaxed without any symmetry constraint until the residual forces are less than 0.01 eV/\AA.

%fig01
\begin{figure}[tbp]
\includegraphics[width=0.6\textwidth]{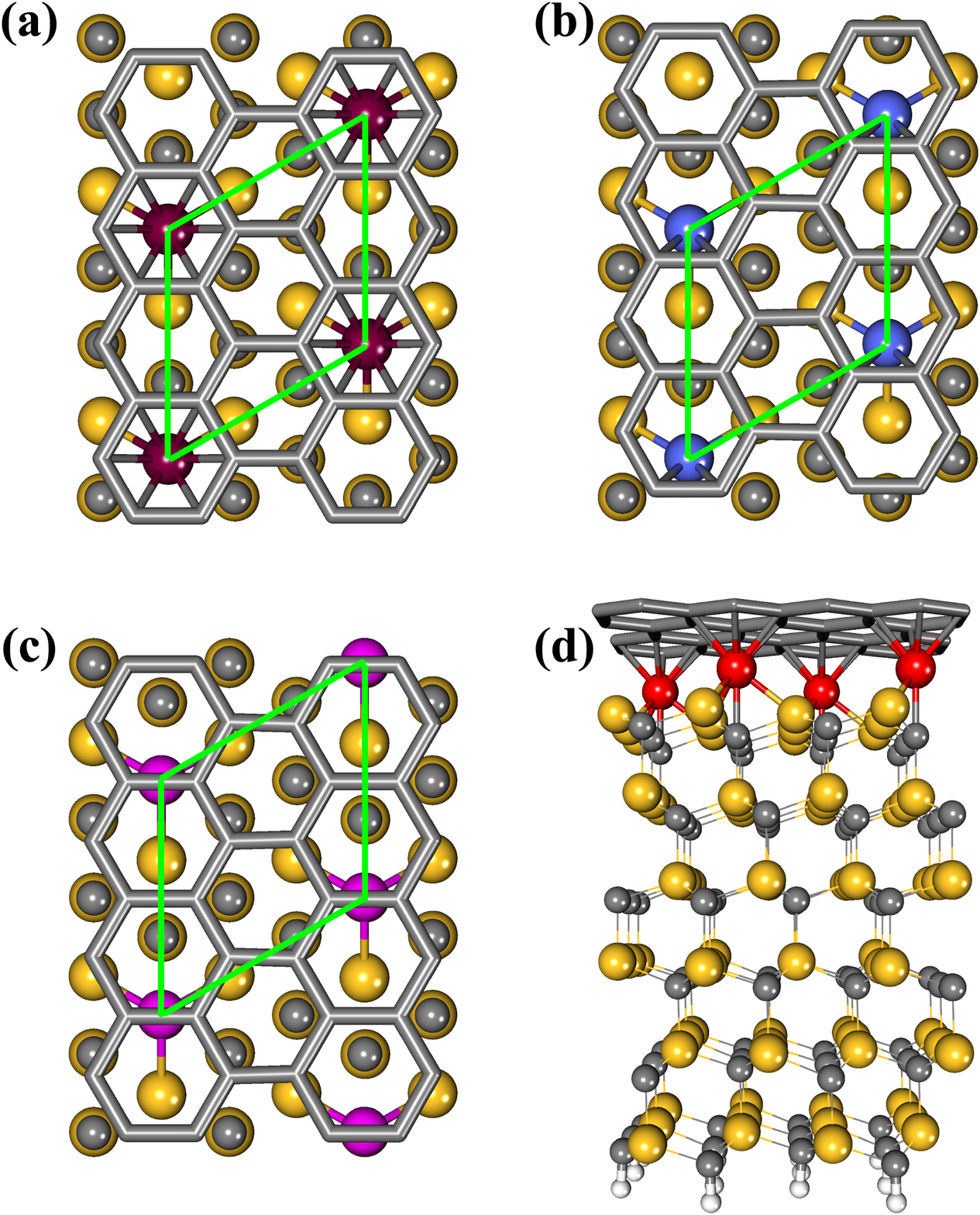}
\caption{\label{fig:fig1}  (Color online) Top view of the optimized geometries of the G/i-TM/SiC systems, (a) Sc, Ti, V, Cr, Mn and Fe, (b) Co, and (c) Ni. (d) Side view of the geometry corresponding to (a). Yellow, gray, and white balls are Si, C, and H, respectively, while others are TM elements. For clarity, graphene is shown as hexagonal mesh. Rhombus denotes the supercell.}
\end{figure}

Quantum transport calculations are conducted using the ATK package \cite{ATK}, which implements nonequilibrium Green's-function formalism within density-functional theory (DFT). The $E_b$ of the graphene with the TM/SiC substrate (per carbon atom) is defined as: $E_{b}=(E_{{\rm G}}+E_{\rm TM/SiC}-E_{{\rm G/i-TM/SiC}})/8$, where $E_{{\rm G}}$, $E_{\rm TM/SiC}$ and $E_{{\rm G/i-TM/SiC}}$ are the total energies of isolated graphene, TM/SiC, and the combined G/i-TM/SiC systems, respectively. The positive value here means that the process is exothermic.

In all the G/i-TM/SiC systems, it was found that the TMs bind to three surface Si atoms with a local $C_{3v}$ symmetry. In contrast, the positions of the TMs with respect to the hexagon of graphene vary. For example, Sc, Ti, V, Cr, Mn, and Fe reside directly below the center of the hexagon [see Fig. 1(a)], Co is about half way away from the center [see Fig. 1(b)], whereas Ni is on the edge directly below a C-C bond [see Fig. 1(c)]. These are distinctly different from the adsorption of TMs on graphene, where the low-energy site is always the center of the hexagon \cite{Valencia}. As compared to G/SiC, the corrugation of graphene is significantly reduced to practically negligible for Sc, Ti, V, Cr, Mn, and Fe [see Fig. 1(d)], and to only $\sim$0.1 \AA\ for Co and Ni. The separation between the TM layer and graphene is $\sim$1.6 \AA\ in most cases except for Ni where the separation is 1.88 \AA.

We characterize G/i-TM/SiC as strongly-bound graphene systems: local density approximation (LDA) calculation reveals that $E_b$ for graphene on TM/SiC $\ge$ 0.24 eV/C \cite{note}, which is at least three times the calculated adsorption energy of graphene on Pd(111) using the same approach \cite{Giovannetti}. Graphene on Mn/SiC has the largest $E_b$ = 0.51 eV/C. This value is even larger than that of graphene on SiC of 0.36 eV/C \cite{Mattausch}. It is known that LDA typically overestimates $E_b$, while generalized gradient approximation (GGA) typically underestimates $E_b$, although the relative energies will not change significantly. Our GGA \cite{pbe} value for G/i-Mn/SiC is $E_b$ = 0.35 eV/C, which can be contrasted to the GGA result for van der Waals binding $-$ typically less than 0.1 eV/C. Therefore, the binding of graphene to TM/SiC is indeed reasonably strong. To experimentally realize the G/i-TM/SiC structures, one approach is to make use of the fact that metal atoms can diffuse through the honeycomb of epitaxial graphene \cite{Virojanadara,jpc,Gierz}. Another approach is to deposit a sub-monolayer TM on the SiC surfaces before covering them with graphene. Upon increasing the temperature, the TM atoms are expected to diffuse to the lowest-energy positions shown in Fig. 1.

%fig02
\begin{figure}[tbp]
\includegraphics[width=0.8\textwidth]{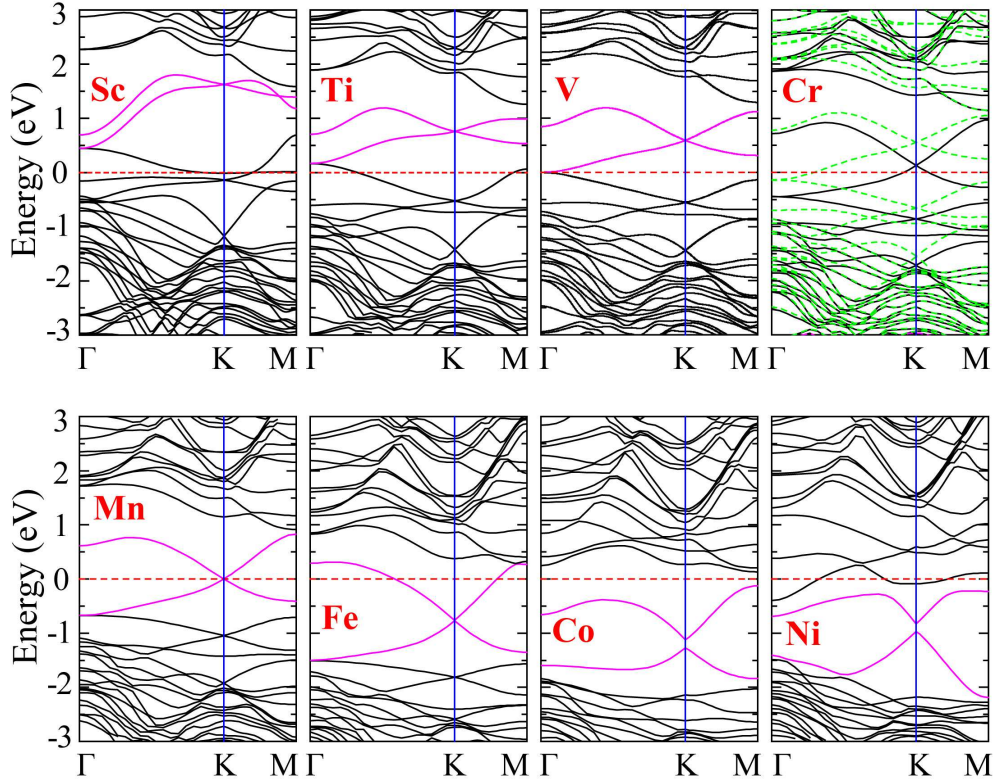}
\caption{\label{fig:fig2} (Color online) Band structures of the G/i-TM/SiC systems. Bands having a linear (or nearly linear) dispersion around the $K$ point and near the $E_F$ are highlighted by magenta except for spin-polarized G/i-Cr/SiC for which black solid lines denote majority-spin states and green dashed lines denote minority-spin states. Fermi levels are indicated by the red horizontal dashed lines at energy zero.}
\end{figure}

Figure 2 shows the calculated band structures for G/i-TM/SiC. The main feature of the bands can be characterized as the existence of two linear bands in the vicinity of $E_F$, highlighted by magenta lines. For Sc, Ti, V, Cr, Mn and Fe, the two bands cross at the $K$ point, forming a Dirac cone. For Co and Ni, the Dirac cone splits, resulting in a small gap of approximately 0.15 eV. Going from Sc to Ni, the position of the Dirac point shifts from above to below $E_F$, indicating a conduction-type change from p to n. G/i-Mn/SiC is a special case where the Dirac point lies exactly at $E_F$. G/i-Co/SiC is also unique in that it has a direct gap of 0.26 eV at the $M$ point. Although both TM-doped graphene \cite{Valencia} and TM-doped SiC \cite{Miao} are magnetic systems, most of the systems we studied are non-magnetic. G/i-Cr/SiC is the only exception. A fully spin-polarized Dirac cone may exhibit negative refraction for one spin, while maintaining positive refraction for the other spin. When an electron beam travels through an interface between regions with different types of carriers, it can be deflected and, depending on the shape of the interface, focused. Combining these two effects may give rise to focused spin-polarized electron beams (spin lens) \cite{Moghaddam}. G/i-Cr/SiC would be such a system for experimental realization of spin lens.

At first glance, the Dirac cone seems to originate from graphene. Two pieces of evidence, however, indicate that this cannot be the case. First, metals have significantly smaller workfunctions than graphene. From our calculations, the workfunctions of TM/SiC are all less than 3.3 eV, in contrast to $\sim$4.6 eV for graphene \cite{Yan}. As a result, electrons are always transferred from TM/SiC to graphene. Charge density analysis reveals that the amount of charge transferred in G/i-TM/SiC is typically 0.5 e/TM. Thus, a Dirac cone that is primarily attributed to graphene cannot be above the $E_F$, as in the case of Sc, Ti, V and Cr. This is qualitatively different from decoupled graphene by intercalating H \cite{Riedl}, Ge \cite{Emtsevprb}, Li \cite{Virojanadara,jpc}, or Au \cite{Gierz} between graphene and SiC. Second, it has been shown that a strong coupling between graphene and TM may destroy the linear dispersion of graphene \cite{Giovannetti}. Therefore, there must be some yet unknown mechanism responsible for the observation of the Dirac cones in Fig. 2.

%fig03
\begin{figure}[tbp]
\includegraphics[width=0.7\textwidth]{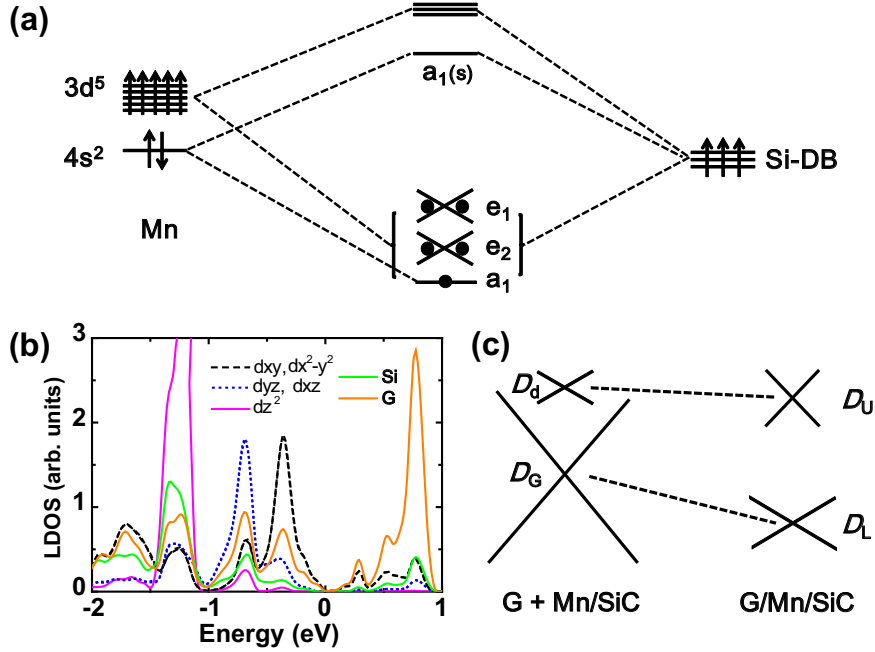}
\caption{\label{fig:fig3} (Color online) (a) Schematic drawing of the hybridization between Mn orbitals and the three dangling-bond orbitals of surface Si in Mn/SiC. Each arrow denotes one electron with up or down spin, while each black dot denotes two (spin-degenerate) electrons. (b) Local density of states (LDOS) of G/i-Mn/SiC. Zero is the Fermi level. (c) Schematic drawing of the hybridization between the Dirac cone of Mn/SiC and that of graphene: left is before and right is after.}
\end{figure}

As a prototypical example, let us examine G/i-Mn/SiC more closely. Previous work has shown that Mn binds \emph{weakly} to graphene due to the high Mn $4s$-to-$3d$ electron promotion energy \cite{Valencia}. Therefore, it has to be the SiC substrate that alters the chemistry of Mn to increase its binding to graphene. Figure 3(a) shows schematically the interactions between Mn and Si
dangling-bond orbitals, derived from the calculated local density of states (LDOS) for Mn/SiC. Note that in the calculation, Mn is kept at the position before graphene removal. The interactions lead to six hybrid states, $a_1$, $e_2$(2), $e_1$(2), and $a_1$(s), respectively, with predominantly metal character: the $a_1$ singlet consists of Mn $4s$ and $d_{\rm z^2}$ character; the $e_2$ doublet consists of Mn $d_{\rm xy}$ and $d_{\rm x^2-y^2}$ character; the $e_1$ doublet consists of Mn $d_{\rm xz}$ and \emph d$_{\rm yz}$ character; and the $a_1$(s) singlet consists of Mn $4s$ character. Only the $a_1$ and $e_2$ states have noticeable contributions from Si dangling bond orbitals. Ten electrons, seven from Mn and three from Si dangling bonds, fully occupy the $a_1$, $e_2$ and $e_1$ states, respectively, leaving the high-lying $a_1$(s) state empty. Critically important is that both the $d$-band $e_2$ and $e_1$ states become Dirac cones at the $K$ point of the Brillouin zone, as
schematically shown in Fig. 3(a). Therefore, the presence of SiC gives rise to two important effects: (1) forcing the donation of Mn $4s$ electrons to Mn $3d$ orbitals and (2) mediating the formation of (relatively-flat) $d$-band Dirac cones.

Next, we consider the hybridization between graphene and Mn/SiC. Figure 3(b) plots the LDOS for G/i-Mn/SiC, which shows clearly the dominant Mn $d$-character of the Dirac cones in Fig. 2. As mentioned earlier, before hybridization both graphene and Mn/SiC have their Dirac cones. Because of the workfunction difference, the Dirac point of Mn/SiC [denoted as D$_d$ in Fig. 3(c)] is above that of graphene [denoted as D$_G$]. Upon hybridization, the (originally nearly flat) Dirac cone of Mn/SiC is significantly broadened, whereas that of graphene is significantly narrowed, as deduced from the calculated band structure of G/i-Mn/SiC in Fig. 2. To qualitatively understand the effects of hybridization, we constructed a 4 $\times$ 4 matrix to mimic the interaction between two Dirac cones:
\begin{equation}
\left(
\begin{array}{cccc}
 -\epsilon+c_{\rm G}k & 0 & a & b \\
 0 & -\epsilon-c_{\rm G}k & b & a \\
 a^\ast & b^\ast & c_{\rm TM}k & 0 \\
 b^\ast & a^\ast & 0 & -c_{\rm TM}k
\end{array}
\right)
\end{equation}
where $\hbar$ is set to 1, $k$ is the wavevector measured from the $K$ point, $\epsilon >$ 0 is the energy separation between the two Dirac points, $c_{\rm G}$ and $c_{\rm TM}$ are the speeds of Dirac fermions for graphene and TM, respectively, and $a$ and $b$ are coupling constants between states of the same and opposite directions of velocity. While the matrix can be numerically diagonalized, it is instructive to examine the solutions when $a$ = 0 and separately when $b$ = 0.

When $a$ = 0 and $|b|\ll\epsilon$, the hybridization results in a uniform reduction in the speeds of graphene and TM by a factor $|b/\epsilon|^2$($c_{\rm G}$ + $c_{\rm TM}$). When $b= 0$  and $|a|\ll\epsilon$, on the other hand, the coupling results in a reduction in the speed of graphene by $|a/\epsilon|^2$($c_{\rm G} - c_{\rm TM}$) but an increase in the speed of TM by $|a/\epsilon|^2$($c_{\rm G} - c_{\rm TM}$), as we have empirically observed. Our model fitting to the calculated results, without the small $a$ assumption, shows
that the dispersion for Mn in Fig. 2 corresponds roughly to $b = 0$ and $a$ = 3.5 eV for $\epsilon$ = 1.5 eV, $c_{\rm G}$ = 1.05 $\times 10^6$ m/s, and $c_{\rm TM} = 10^5$ m/s. We note that the actual coupling may involve multi-bands and is hence more complex. Our fitting here is only to provide a physical picture. For a high crystal symmetry as the one in Fig. 1(a), it is conceivable that the coupling between waves of opposite traveling directions, namely, $b$, is small. However, for a lower crystal symmetry as in the case of Co in Fig. 1(b) or Ni in Fig. 1(c), however, this is highly unlikely. When neither $a$ nor $b$ is zero, our model suggests a gap at the Dirac point:
$E_g = (\sqrt{\epsilon^2+4(a+b)^2}-\sqrt {\epsilon^2+4(a-b)^2})/2\approx 4ab/\sqrt{\epsilon^2+4a^2}$ for $b \ll a,\epsilon$. The presence of a gap is consistent with the results of direct calculations
for Co and Ni, showing a gap of about 0.15 eV. Graphene has about 8\% lattice mismatch to SiC \cite{Mattausch}. Using a 6$\sqrt 3\times 6\sqrt 3R$30$^\circ$ surface cell, which is nearly
lattice matched to a $13 \times 13$ graphene cell, we find a gap of 0.16 eV for G/i-Mn/SiC. The gap opening here can be attributed to an increased lattice corrugation similar to that of
Co and Ni. Despite the gap opening, our fitting also reveals that the linear dispersion remains intact as long as the wavevector $k$ (measured from the $K$ point) is larger than 0.01 $\pi/a$. These small-gap G/i-TM/SiC systems may be used as planar linear-band semiconductors.

%fig04
\begin{figure}[tbp]
\includegraphics[width=0.7\textwidth]{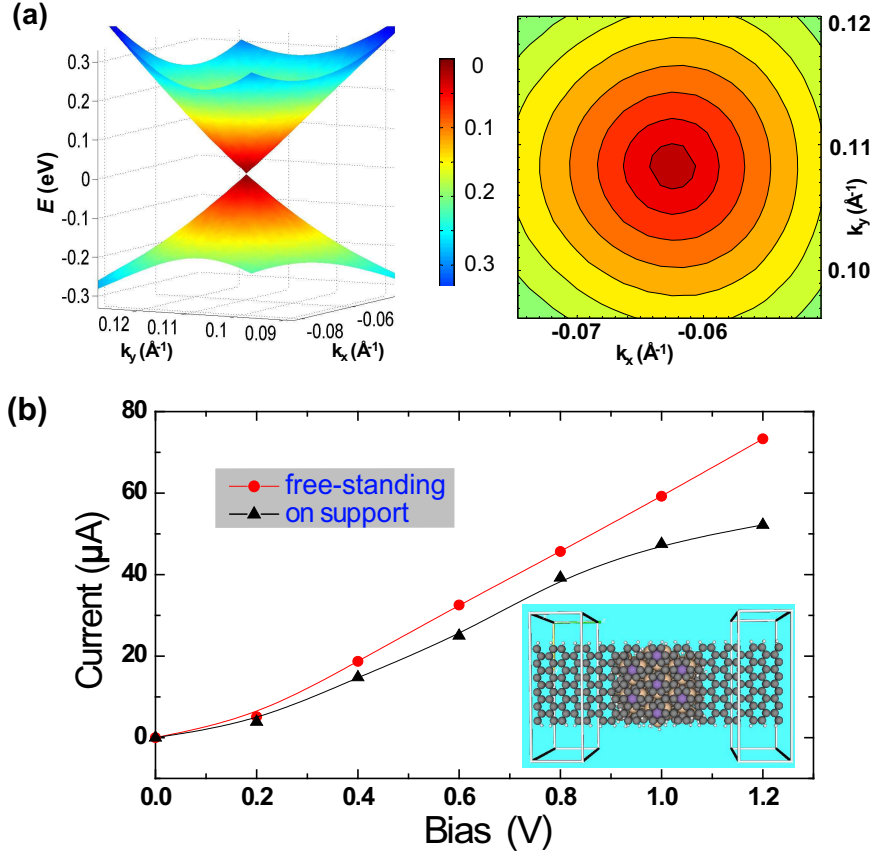}
\caption{\label{fig:fig4}(Color online) (a) Band structure around the Dirac point for G/i-Mn/SiC. Left: three-dimensional view and right: constant energy contours. Color represents vertical distance from zero energy surface. (b) The \emph{I}-\emph{V} curves for 11-AGNR ($=$ armchair graphene nanoribbon with 11 dimer rows) with and without the Mn/SiC support. Inset shows the atomic structure for transport calculations.}
\end{figure}

Due to the symmetry of the $d$ orbitals, one may expect the $d$-state-derived Dirac cone to be highly anisotropic \cite{Park}. If so, group velocity may be significantly altered, at least in certain directions. The left panel of Fig. 4(a) shows a three-dimensional plot of the Dirac cone for G/i-Mn/SiC. It reveals that the Dirac cone is as isotropic as that in graphene with a Fermi velocity of 0.5 $\times$ 10$^6$ m/s, which is almost half that in free-standing graphene. A linear dispersion is maintained for energies within $\pm$0.3 eV from the Dirac point, which is more than half of that for graphene, $\pm$0.5 eV \cite{Wallace}. The constant-energy contours of the valence band around the Dirac point is shown in the right panel of Fig. 4(a). It confirms that all the contours are concentric circles as are those in graphene. Figure 4(b) shows the calculated current-voltage (\emph{I-V}) curves. It can be seen that the \emph{I-V} characteristic of G/i-Mn/SiC under a moderate bias of less than $\sim$0.6 V is also similar to that of graphene. Thus, a free-standing graphene-like transport can also be achieved on such strongly-bound systems.

In summary, first-principles calculations reveal Dirac spectra in strongly-bound G/i-TM/SiC systems. The SiC substrate modifies the electronic structure of the TM $d$ states to foster a strong coupling between TM and graphene. This also leads to a migration of the Dirac cone at the Fermi level from graphene to TM, which can be qualitatively understood via a simple model Hamiltonian. In the case of G/i-Mn/SiC, the Fermi level is at the Dirac point. Despite the $d$ character of the states, the Dirac fermions behave the same way as those in graphene except for a half Fermi velocity. For other TM, our study also reveals interesting physical phenomena such as symmetry breaking, gap opening and spin-polarized Dirac cone near $E_F$. These results point to alternative and perhaps more realizable directions for graphene-based electronics and spintronics.

We acknowledge the support of the Ministry of Science and Technology of China (Grant Nos. 2011CB921901 and 2011CB606405), and the National Natural Science Foundation of China (Grant Nos. 11074139,  11174167 and 11104155). SBZ acknowledge the supports by the National Nuclear Security Administration, Office of Nuclear Nonproliferation Research and Engineering (NA-22), of U.S. DOE and DOE/BES under Contract No. DE-SC0002623.


\begin{references}
\bibitem{note} The exact values are 0.24, 0.36, 0.40, 0.43, 0.51, 0.44, 0.34 and 0.26 eV/C for Sc to Ni intercalated graphene/SiC(0001), respectively.
\end{references}
\end{document}